\documentclass[reprint,twocolumn,superscriptaddress,amsmath,amssymb,showpacs,prl]{revtex4-1}
\usepackage{hyperref}
\pdfoutput=1 

\usepackage{dcolumn}
\usepackage{epsfig}
\usepackage{graphics}
\usepackage[latin1]{inputenc} 
\usepackage[T1]{fontenc}
\usepackage{bm}
\usepackage{hyperref}
\newcommand{\ddst}{false}

\begin{document}

\title{Creep of Bulk C--S--H: Insights from Molecular Dynamics Simulations}

 \author{Mathieu Bauchy}
 \email[Contact: ]{bauchy@ucla.edu}
 \homepage[\\Homepage: ]{http://mathieu.bauchy.com}
 \affiliation{Physics of AmoRphous and Inorganic Solids Laboratory (PARISlab), Department of Civil and Environmental Engineering, University of California, Los Angeles, CA 90095, United States}
 \author{Enrico Masoero}
 \affiliation{School of Civil Engineering and Geosciences, Newcastle University, Newcastle upon Tyne, NE1 7RU, United Kingdom}
 \author{Franz-Joseph Ulm}
 \affiliation{Concrete Sustainability Hub, Department of Civil and Environmental Engineering, Massachusetts Institute of Technology, 77 Massachusetts Avenue, Cambridge, MA 02139, United States}
 \affiliation{MIT-CNRS joint laboratory at Massachusetts Institute of Technology, 77 Massachusetts Avenue, Cambridge, MA 02139, United States}
 \author{Roland Pellenq}
 \affiliation{Concrete Sustainability Hub, Department of Civil and Environmental Engineering, Massachusetts Institute of Technology, 77 Massachusetts Avenue, Cambridge, MA 02139, United States}
 \affiliation{MIT-CNRS joint laboratory at Massachusetts Institute of Technology, 77 Massachusetts Avenue, Cambridge, MA 02139, United States}
 \affiliation{Centre Interdisciplinaire des Nanosciences de Marseille, CNRS and Aix-Marseille University, Campus de Luminy, Marseille, 13288 Cedex 09, France}

\date{\today}


\begin{abstract}
Understanding the physical origin of creep in calcium--silicate--hydrate (C--S--H) is of primary importance, both for fundamental and practical interest. Here, we present a new method, based on molecular dynamics simulation, allowing us to simulate the long-term visco-elastic deformations of C--S--H. Under a given shear stress, C--S--H features a gradually increasing shear strain, which follows a logarithmic law. The computed creep modulus is found to be independent of the shear stress applied and is in excellent agreement with nanoindentation measurements, as extrapolated to zero porosity.
\end{abstract}

\maketitle

\section{Introduction}

Creep is a major limitation of concrete. Indeed, it has been suggested that creep deformations are logarithmic, that is, virtually infinite and without asymptotic bound, which raises safety issues \cite{bazant_excessive_2011}. The creep of concrete is generally thought to be mainly caused by the viscoelastic and viscoplastic behavior of the cement hydrates \cite{acker_swelling_2004}. While secondary cementitious phases can show viscoelastic behavior \cite{nguyen_microindentation_2012}, the rate and extent of viscoelastic deformations of such phases is far less significant than that calcium--silicate--hydrate (C--S--H), the binding phase of the cement paste \cite{acker_swelling_2004}. As such, understanding the physical mechanism of the creep of C--S--H is of primary importance.

Despite the prevalence of concrete in the built environment, the molecular structure of C--S--H has just recently been proposed \cite{pellenq_realistic_2009, abdolhosseini_qomi_combinatorial_2014}, which makes it possible to investigate its mechanical properties at the atomic scale. Here, relying on the newly available model, we present a new methodology allowing us to simulate the long-term creep deformation of bulk C--S--H (at zero porosity, i.e., at the scale of the grains). Results show an excellent agreement with nanoindentation measurements \cite{vandamme_nanogranular_2010}.

\section{Simulation details}

To describe the disordered molecular structure of C--S--H, Pellenq et al. \cite{pellenq_realistic_2009} proposed a realistic model for C--S--H with the stoichiometry of (CaO)$_{1.65}$(SiO$_2$)(H$_2$O)$_{1.73}$. We generated the C--S--H model by introducing defects in an 11 \AA\ tobermorite \cite{hamid_crystal-structure_1981} configuration, following a combinatorial procedure. Whereas the Ca/Si ratio in 11 \AA\ tobermorite is 1, this ratio is increased to 1.71 in the present C--S--H model, through randomly introducing defects in the silicate chains, which provides sites for adsorption of extra water molecules. The ReaxFF potential \cite{manzano_confined_2012}, a reactive potential, was then used to account for the reaction of the interlayer water with the defective calcium--silicate sheets. More details on the preparation of the model and its experimental validation can be found in Ref. \cite{abdolhosseini_qomi_combinatorial_2014} and in previous works \cite{bauchy_order_2014,pellenq_realistic_2009,abdolhosseini_qomi_applying_2013,abdolhosseini_qomi_anomalous_2014,bauchy_nanoscale_2014,bauchy_fracture_2014,bauchy_is_2014,bauchy_topological_2014, bauchy_rigidity_2015}.

We simulated the previously presented C--S--H model, made of 501 atoms, by molecular dynamics (MD) using the LAMMPS package \cite{plimpton_fast_1995}. To this end, we used the REAXFF potential \cite{manzano_confined_2012} with a time step of 0.25fs. Prior to the application of any stress, the system is fully relaxed to zero pressure at 300K.

\section{Results}

\begin{figure}
\begin{center}
\includegraphics*[width=\linewidth, keepaspectratio=true, draft=\ddst]{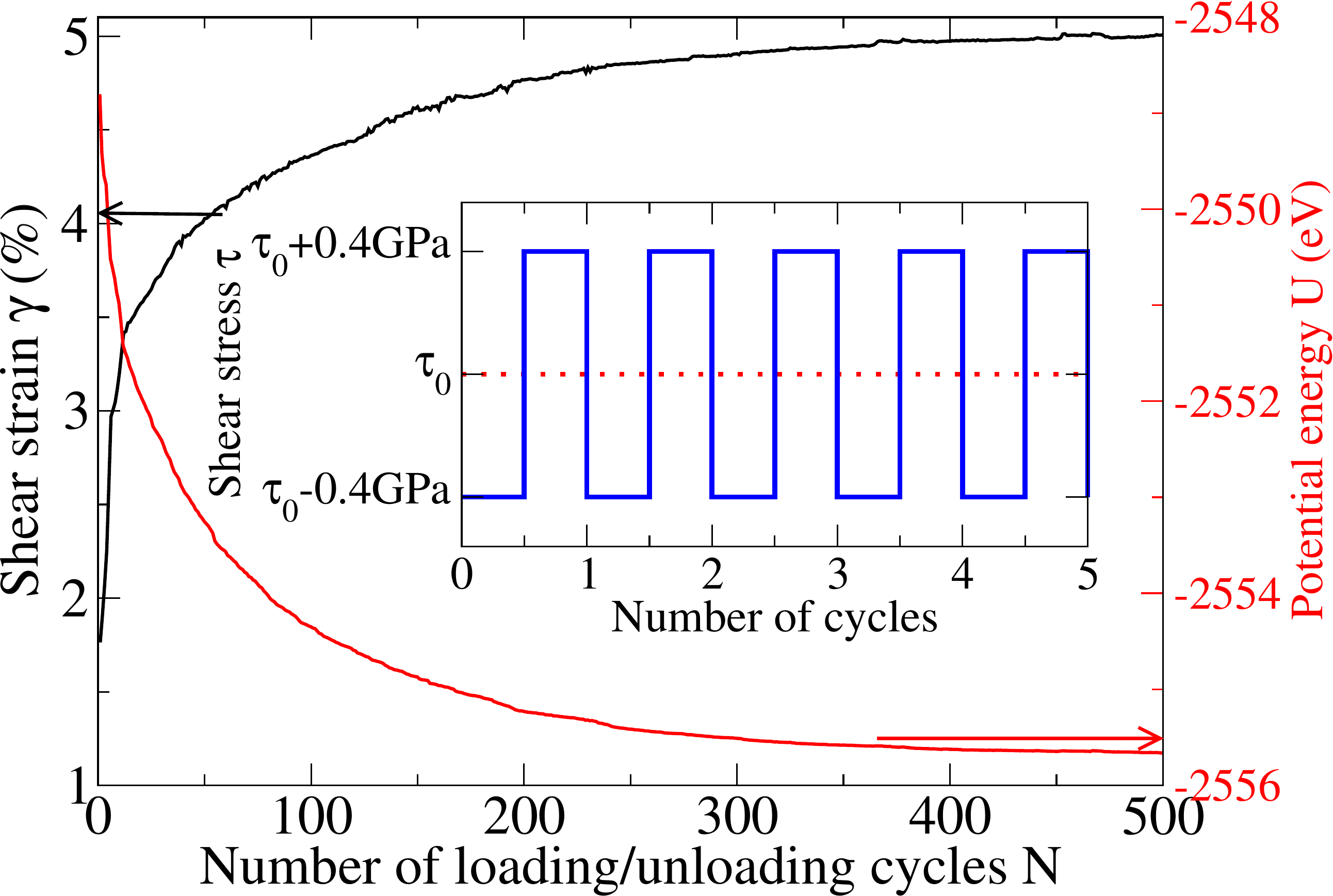}
\caption{\label{fig:method} Shear strain and potential energy with respect to the number of loading/unloading cycles. The inset shows the shape of the applied shear stress.
}
\end{center}	
\end{figure}

The relaxation of C--S--H, or of other silicate materials, takes place over long periods of time (years), which prevents the use of traditional MD simulations, which are limited to a few nanoseconds. To study the long-term deformations of C--S--H, we applied a method that has recently been introduced to study the relaxation of silicate glasses \cite{yu_stretched_2015}. In this method, starting from an initial atomic configuration of glass, formed by rapid cooling from the liquid state, the system is subjected to small, cyclic perturbations of shear stress $\pm \tau_p$ around zero pressure. For each stress, a minimization of the energy is performed, with the system having the ability to deform (shape and volume) in order to reach the target stress. These small perturbations of stress deform the energy landscape of the glass, allowing the system to jump over energy barriers. Note that the observed relaxation does not depend on the choice of $\tau_p$, provided that this stress remains sub-yield \cite{yu_stretched_2015}. This method mimics the artificial aging observed in granular materials subjected to vibrations \cite{richard_slow_2005}.

Here, in order to study creep deformation, we add to the previous method a constant shear stress $\tau_0$, such that $\tau_p < \tau_0$ (see the inset of Figure \ref{fig:method}). When subjected to shear stresses of different intensities, C--S--H presents a shear strain $\gamma$ that: (1) increases logarithmically with the number of cycles $N$ (Figure \ref{fig:method}) and (2) is proportional to the applied shear stress (see Figure \ref{fig:strain}).

\section{Discussion}

\begin{figure}
\begin{center}
\includegraphics*[width=\linewidth, keepaspectratio=true, draft=\ddst]{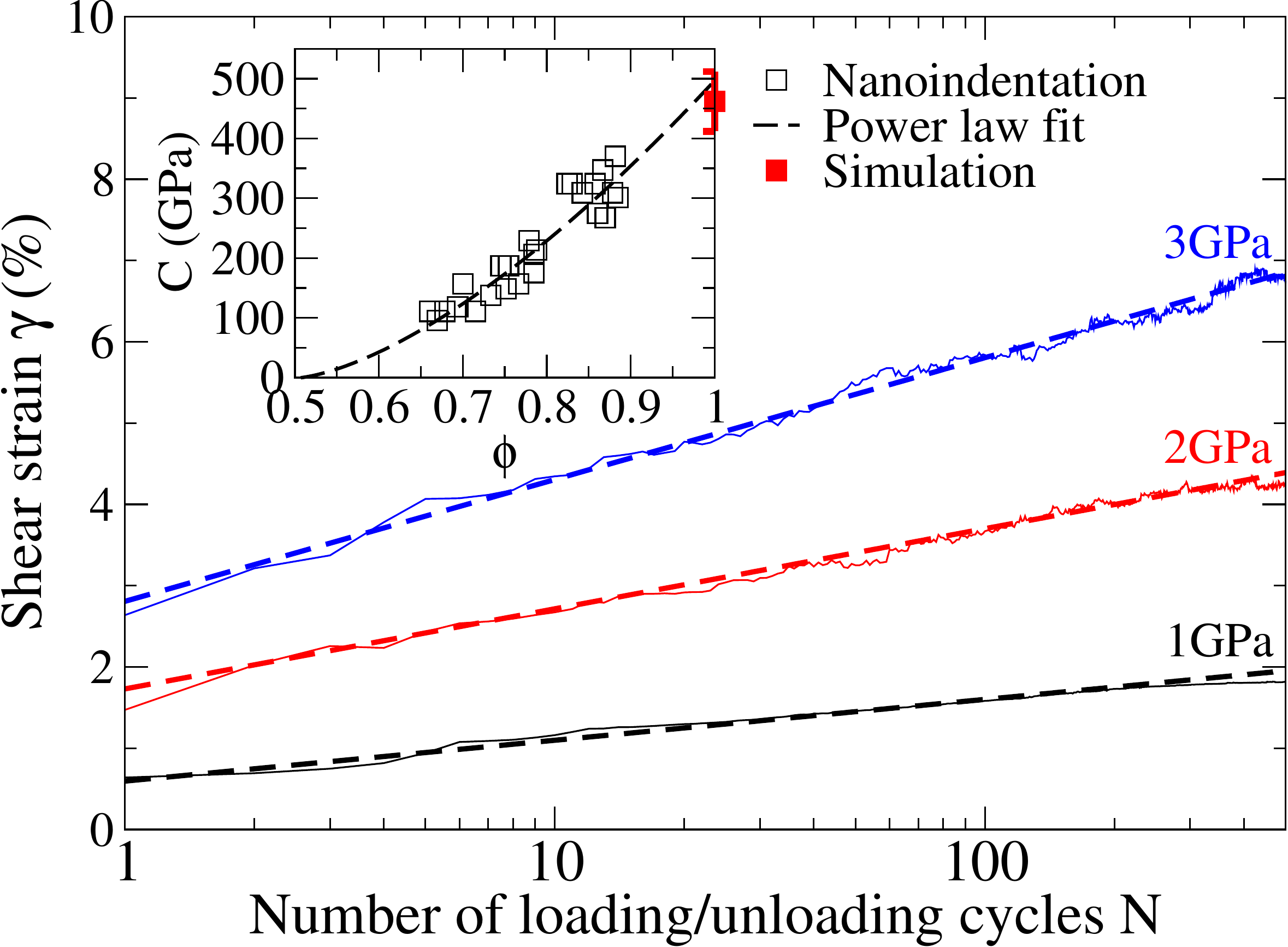}
\caption{\label{fig:strain} Shear strain with respect to the number of loading/unloading cycles, under a constant shear stress of 1, 2, and 3 GPa. The inset shows the creep  modulus $C$ with respect to the packing fraction $\phi$ obtained from nanoindentation \cite{vandamme_nanogranular_2010}, compared with the computed value at $\phi = 0$.
}
\end{center}	
\end{figure}

The creep of bulk C--S--H can then be described by a simple logarithmic law $\gamma (N) = (\tau_0/C) \log(1+N/N_0)$,  where $N_0$ is a constant analogous to a relaxation time and $C$ is the creep modulus. A careful look at the internal energy shows that the height of the energy barriers, through which the system transits across each cycle, remains roughly constant over successive cycles. According to transition state theory, which states that the time needed for a system to jump over an energy barrier $E_A$ is proportional to $\exp(-E_A/kT)$, we can assume that each cycle corresponds to a constant duration $\Delta t$, so that a fictitious time can be defined as $t=N \Delta t$ \cite{masoero_kinetic_2013}.

We note that the computed creep moduli $C$ does not show any significant change with respect to the applied stress $\tau_0$. As such, it appears to be a material property that can directly been compared to nanoindentation results extrapolated to zero porosity \cite{vandamme_nanogranular_2010}. As shown in the inset of Figure \ref{fig:strain}, we observe an excellent agreement, which suggests that the present method offers a realistic description of the creep of C--S--H at the atomic scale. This result also suggests that, within the linear regime (i.e., for sub-yield stresses, when $C$ remain constant), deformations due to cyclic creep and basic creep, with respect to the number of stress cycle or the elapsed time, respectively, should be equivalent.

\section{Conclusion}
We reported a new methodology based on atomistic simulation, allowing us to successfully observe long-term creep deformations of C--S--H. Creep deformations are found to be logarithmic and proportional to the applied shear stress. The computed creep modulus shows an excellent agreement with nanoindentation data, which suggests that the present methodology could be used as a predictive tool to study the creep deformations of alternative binders.

\begin{acknowledgments}
MB acknowledges partial financial support for this research provisioned by the University of California, Los Angeles (UCLA). This work was also supported by Schlumberger under an MIT-Schlumberger research collaboration and by the CSHub at MIT. This work has been partially carried out within the framework of the ICoME2 Labex (ANR-11-LABX-0053) and the A*MIDEX projects (ANR-11-IDEX-0001-02) cofunded by the French program "Investissements d'Avenir" which is managed by the ANR, the French National Research Agency.
\end{acknowledgments}


\end{document}